\documentstyle[twocolumn,aps,prl,overcite]{revtex}
\input epsf

\tighten

\def\be{\begin{equation}}
\def\ee{\end{equation}}
\def\ba{\begin{eqnarray}}
\def\ea{\end{eqnarray}}

\begin{document}

\preprint{\parbox{1.5in}}

\title{
\vskip -.2in
 \hfill    {\normalsize hep-th/0011006 }\\
\vskip 0.1in
 A Self-Tuning Solution of the Cosmological Constant
Problem}

\author{A. Kehagias {}$^a$ and K. Tamvakis{}$^b$}

\address{${}^a$~Physics Department, National Technical University, \\
15773 Zografou, Athens, Greece.}

\address{${}^b$~Physics Department, University of Ioannina\\
45110 Ioannina, Greece.}

\maketitle
\begin{abstract}
We discuss the four-dimensional cosmological constant problem in a
five-dimensional setting. A scalar field coupled to the SM forms
dynamically a smooth brane with four-dimensional Poincar\'e invariance,
independently of SM physics. In this respect,
our solution may be regarded as a self-tuning solution, free of any
singularities and fine-tuning problems.

\end{abstract}

\smallskip




\medskip

Among the long-standing problems in physics, a central one, from a
 theoretical point of view, is the
lack of  any understanding of the smallness of the cosmological
constant\cite{weinberg} . The cosmological constant predicted from
a field theory calculation of the vacuum energy for a cutoff at
the Planck mass $M_{P}$, is of an enormous $\Lambda\sim M_{P}^2$
size, more than 120 orders of magnitude from the  observed bounds
\cite{Bahcall:1999xn} . Supersymmetry does not provide a solution
to the problem, although it would predict a vanishing cosmological
constant in case it were
 an exact symmetry. However, after supersymmetry breaking, a non-zero
cosmological constant $\sim M_{SUSY}^4/M_{P}^2$ has to arises,
still far beyond any realistic value.

Recently, new proposals of relaxing the cosmological constant to
zero or to a really small value have been put forward within the
brane framework. In the brane set up our universe is modeled as a
hypersurface embedded in a higher dimensional continuum. The idea
of a {\textit{``wall-world"}} is not new \cite{TT,Ruba} and in
modern language is realized by {\textit{D-branes}}. These arise in
String Theory and are extended stable objects on which open
strings can end. Standard Model physics is confined on the brane
whereas gravity propagates in the bulk. Nevertheless, bulk
propagation of gravity is in contradiction with the observed fact
of four-dimensional gravitation satisfying an inverse-square
Newton's law. There have been a number of proposals in trying to
isolate a four-dimensional graviton. After succeeding in
localizing a massless graviton on the brane\cite{RSu}, while
massive modes
 introduce a small correction to Newton's law,
 the idea of a brane-resolution of the
 cosmological constant problem\cite{ADS}$^-$ \cite{Krauss}
 has been put forward.
One of the proposals is the {\textit{self-tuning}} one, which
involves  sets of background solutions with 4D Poincar\'e
invariance for arbitrary values of the tension of the Standard
Model brane. However, these solutions,
 existing for a restricted form
of bulk interactions, are singular. The existence of naked
singularities and their connection to fine-tuning makes the
whole proposal questionable\cite{Gu,Wi,FN,Bi}  .
In fact, regarding the bulk scalar employed in the self-tuning
models as a KK scalar, the 5D solution can be lifted\cite{AK} to 6D.
There, one observes that although the metric is regular, there is a global
conical singularity and its resolution requires indeed fine-tuning.

The purpose of this work is to suggest a solution to the
cosmological constant problem in the spirit of self-tuning
proposal without however the drawbacks of singularities and/or
fine-tunings. This is achieved by the dynamical formation of the
4D brane by a bulk five-dimensional scalar. The fact that bulk
scalar fields with non-trivial potential can lead to smooth
backgrounds with the desired localization properties has been
illustrated in reference {\rm 24}. In the present article
 we argue that by introducing the appropriate 5D coupling of the SM to
the brane-forming scalar, not only localization of the SM on a
brane is achieved, but in addition, any contribution to the 4D cosmological
constant related to the SM physics can be neutralized leading to a
4D Poincar\'e invariant background.

Let us now consider the five-dimensional action \ba {\cal S}=\int
d^5x\sqrt{-G}\left(2M^3R-\frac{1}{2}(\partial \phi)^2
-{\cal{L}}_mJ(\phi)\right) \, ,
 \label{act} \ea where ${\cal{L}}_m$
is the five-dimensional matter Lagrangian that gives rise to the
four-dimensional Standard Model Lagragian after localization. It
is in general a functional of gauge, Higgs  and fermion fields
(collectively denoted by $\chi_i$) such that ${\cal L}_m={\cal
L}_m(\chi_i)$. The equations of motion resulting from the action
(\ref{act}) are
\ba
R_{MN}-\frac{1}{2}G_{MN}R&=&\frac{1}{4M^3}\left(
\partial_M\phi\partial_{N} \phi+{\delta {\cal L}_m\over \delta G_{MN}}J(\phi)
\phantom{X\over X^x} \right.  \nonumber  \\ && \left.-
G_{MN}\Big{(}\frac{1}{2} (\partial \phi)^2+U(\phi)\Big{)}
\right)\, ,\label{a1}
\ea

\ba
&&\frac{1}{\sqrt{-G}}\partial_M\left\{\sqrt{-G}G^{MN}\partial_N\phi\right\}
=\frac{\partial U}{\partial \phi}\, , \label{a2}
\\ &&
{\delta\over \delta \chi_i}\Big{(}\sqrt{-G}{\cal
L}_mJ(\phi)\Big{)}=0\, ,
 \label{ll}
\ea
where
\be
U(\phi)=J(\phi){\cal{L}}_m\, ,
\ee
is the potential for the scalar $\phi$.
In what follows a capital index like $M$ stands for $0,1,2,3,4$
while $\mu=0,1,2,3$. The ordinary four coordinates will be
represented by $x^{\mu}$ while for the fifth coordinate we shall
use the symbol $y$. In looking for  four-dimensional Poincar\'e invariant
solutions, the most general form of the metric respecting this
symmetry is
\be
ds^2=e^{2A(y)}\left(-dt^2+dx_1^2+dx_2^2+dx_3^2\right)+dy^2\, ,
\label{met} \ee while we shall consider the scalar $\phi$ to be
only $y$-dependent. In addition, since we are looking for vacuum
solutions, we put all fields in ${\cal L}_m$, as usual,  to zero
except the Higgs scalars $H$. In this case, ${\cal L}_m=-V_{\rm
eff}(H)$
 and
eq.(\ref{ll}) is just the extremality  condition $\partial V_{\rm
eff}/\partial H=0$ of $V_{\rm eff}$. We shall denote the value of
${\cal L}_m$ at the extremum as $V_0$. Then, the equations
eqs.(\ref{a1},\ref{a2}) become equivalent to the pair
\begin{equation}
\frac{1}{2}(\phi')^2-U_0(\phi)=24M^3(A')^2 \, ,\label{e1}
\end{equation}
\begin{equation}
\frac{1}{2}(\phi')^2+U_0(\phi)=-12M^3A^{''}-24M^3(A')^2\, ,
\label{e2}
\end{equation}
with
\be
U_0(\phi)\equiv V_0J(\phi) \label{u0} \, .
\ee
All Standard Model physics
 is contained in the parameter $V_0$. The
system of equations (\ref{e1},\ref{e2}) will,
presumably, produce as a solution the pair $A,\,\phi$ as functions of $y$.
Equivalently, we can consider the pair of pair $A,\,\phi'$
{\textit{as functions of $\phi$}}. Then, since
$A'=\phi'\frac{\partial A}{\partial \phi}$, we can always write
\be
A'=-\frac{W(\phi)}{12M^3}\,  , \label{AA}
\ee
where $W(\phi)$ is a function of $\phi$ defined by this equation
and called {\textit{superpotential}}\cite{ST} . Subtracting equations
(\ref{e1},\ref{e2}) leads us to an expression of the potential in
terms of the superpotential
\be
U_0(\phi)=\frac{1}{2}\left(\frac{\partial W}{\partial
\phi}\right)^2
-\frac{W^2}{6M^3}\, , \label{u}
\ee
 The sum of eqs.(\ref{e1},\ref{e2}),
gives us
\be
\phi'={\partial W\over \partial\phi}\, . \label{ff} \ee
Note that the above form of the potential is not an assumption
but rather a necessary condition for smooth  solutions
to exist.
This restriction ceases to hold for more than one fields.
Thus, in our case, finding an appropriate solution
is translated into choosing an appropriate superpotential $W$.
 Taking the trial choice
$$W=\gamma\sin(\beta \phi)\, ,$$ we obtain the sine-Gordon form,
also employed elswhere\cite{Gr,MM} of the potential\footnote{Note the
symmetry of the potential $\phi\rightarrow \phi +2\pi n/\beta$ for
integer $n$. A potential of this form is expected to arise in
theories of antisymmetric tensor fields\cite{Q}.}
\ba
U_0=
{\gamma^2\beta^2\over 2}\left(1-g^2\sin^2(\beta \phi)\right)\, .
\label{uuu} \ea
 We have introduced
 \be
 g^2=1+\frac{1}{3M^3\beta^2}\, . \label{b}
 \ee
  It can easily be seen
that the potential (\ref{uuu}) is of  the form (\ref{u0}) with $V_0=\frac{1}{2}\gamma^2\beta^2$ and $V_0>0$. We could
alternatively have started with a potential
$U_0(\phi)=V_0\left(1-g^2\sin^2(\beta\phi)\right)$. Then, we would
have the superpotential form (\ref{u}) of the potential, and,
therefore, a solution, only when $\beta $ and $g^2$ would
 be restricted by (\ref{b}). It should be stressed however that eq.(\ref{b})
{\textit{ is totally independent of the Standard Model Physics
represented by $V_0$. It is not, therefore, a fine-tuning but
a restriction in a subspace of the
bulk-parameter space of $\beta$, $g^2$}}.

Choosing $V_0$ and $g^2$ as our parameters, we can proceed to
solve for $\phi(y)$ and $A(y)$.
The scalar field solution can be obtained from eq.(\ref{ff}).
It has the known kink-like form
\be
\phi=\frac{2}{\beta}\arctan\Big{(}\tanh(ay/2)\Big{)} \, , \label{sgb}
\label{f11}
\ee
where $a^2\equiv 2V_0\beta^2=\frac{2V_0}{3M^3(g^2-1)}$. Note the restriction $g^2>1$. Similarly, we can solve for $A(y)$
eq.(\ref{AA}) by using
(\ref{f11}) and we obtain the {\textit{warp function}}
\be
A(y)=-\frac{(g^2-1)}{8}\ln\cosh^2(ay)\, .\label{wf} \ee The
background geometry is then described by the metric\footnote{The
four-dimensional Planck mass has a finite value
$$M_P^2=4M^3\xi^{-1}\sqrt{\pi}
\frac{\Gamma\left((g^2+3)/4)\right)}{\Gamma\left((g^2+1)/4\right)}$$}
\be
ds^2=\left\{\cosh(ay)\right\}^{(1-
g^2)/ 2}\eta_{\mu\nu}dx^\mu dx^\nu
+dy^2\, .\label{mt}
\ee

 The calculation of  all curvature invariants  like $R$ and
$R_{MN}R^{MN}$, demonstrates that the geometry is nowhere singular
for $g^2>1$. For every pair of values $V_0,\,g^2>1$ there exists a
smooth geometry given by (\ref{wf}) and a smooth bounce-like field
configuration given by (\ref{sgb}). Quantum SM effects
may change the value of $V_0$ but there will always be a solution
corresponding to the modified value without any effect on the bulk
parameter $g^2$.

We give below in fig.1 a graph of the solution for some special
values of the parameters.
 \vskip .5in

\centerline{\epsfxsize=200pt\epsfysize=200pt\epsfbox{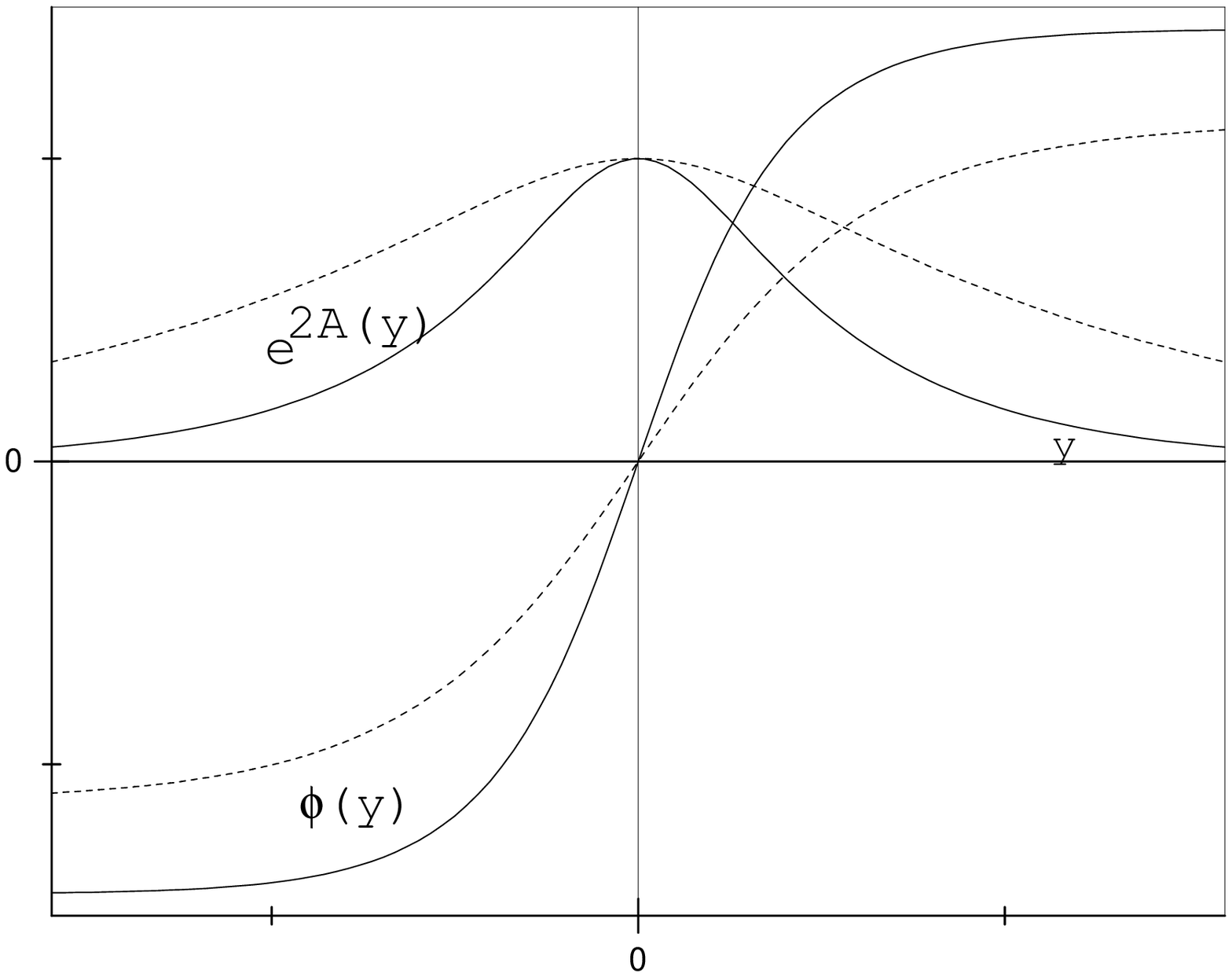}}
\medskip\noindent
{\small Fig.~1: The functions $e^{A(y)},\ \phi(y)$ for $V_0=1\, ,
\beta=1.1$ (solid lines) and $V=0.25\, , \beta=1.4$ (dashed
lines).}

\vskip .1in

There is a {\textit{brane-limit}} to the smooth solution found above defined as
$g^2\rightarrow 1\,\,\,\,,\,\,\,2V_0(g^2-1)/3M^3
\equiv \xi^2<\infty $. In this limit the warp function takes the
 Randall Sundrum\cite{RSu} form $A(y)\rightarrow -\frac{\xi}{4}|y|$.
 If we substitute our
solution to the bulk matter action in this limit, we
obtain
 $$\frac{1}{2}(\phi')^2+U_0(\phi) \rightarrow -\frac{3}{2}M^3\xi^2+12M^3\xi\delta(y)\,.$$
$\xi$ is by definition positive. Thus, in this limit
the sine-Gordon bounce behaves as a brane of tension $12M^3\xi$
placed at $y=0$.
It should be remarked that both the bulk kinetic term $(\phi')^2/2$
and the potential $U_0(\phi)$ contribute equally to the brane term.
 Localization
of SM fields will not be automatic. Massless scalars and chiral
fermions in the SM Lagrangian are going to be localized by the
$e^{2A}$ warp factor,
 while for gauge bosons a separate localization
mechanism will be required\cite{KT}.

The coupling function $J(\phi)$ employed above, changes sign away from the
{\textit{brane}} at $y=0$. Although this is not
something that one should necessarily worry about, since SM fields
will
 somehow get to be localized, it is important to show that we can replace the coupling function
with one of constant sign. We can consider the superpotential
\begin{equation}
W=\frac{\beta}{4\alpha^2}\left(2\alpha\phi+\sin(2\alpha \phi)\right)\, ,
\end{equation}
that leads to $\phi=\frac{1}{\alpha}\arctan(\beta y)+2\pi
n/\alpha$, where $n$ is an arbitrary integer, and to the perfectly
localizing warp function
\begin{equation}A(y)=-\mu y\arctan(\beta
y)\, ,
\end{equation}
with $\mu=\beta/24M^3\alpha^2$. The
potential $U_0(\phi)$ resulting from $W(\phi)$ is
\begin{equation}\frac{\beta^2}{8\alpha^2}
\left\{\left(1+\cos(2\alpha \phi)\right)^2-\frac{1}{12M^3\alpha^2}
\left(2\alpha \phi+\sin(2\alpha \phi)\right)^2\right\}\, .
\label{ppp}
\end{equation}
Note that this potential does not posses the symmetry
$\phi\rightarrow \phi+n\pi/\alpha$ present in the previous
example. Expressing $U_0$ as a function of $y$ shows that for a
sufficiently large value of $n$ the potential will be always
negative and thus, the SM Lagrangian has the correct sign. A
detailed study of the model with the potential of eq.(\ref{ppp}) will
be given elsewhere \cite{KT2}.

Although we have constructed a flat solution with four-dimensional
Poincar\'e invariance, we have not finished yet. We have to make
sure that a localized four-dimensional massless graviton exists.
After all, the cosmological constant is a problem as long as
gravity is present. For this, let us consider a perturbation
around the previously described solution of the form
\begin{equation}\delta G_{MN}=\delta_M^{\mu}\delta_N^{\nu}h_{\mu\nu}(x,y)
\,\,\,\,,\,\,\,\,\delta\phi=0\, .
\end{equation}
$h_{\mu\nu}$ represents the graviton in the axial gauge defined by
the constraint $h_{5M}=0$.
We are interested on the transverse modes and, therefore, we shall assume
$ h_{\mu}^{\mu}=\partial_{\mu}h^{\mu\nu}=0$. The equations give, to first order in $h_{\mu\nu}$,
$$\left\{-\frac{1}{2}\frac{\partial^2}{\partial y^2}+
\left(A^{''}+2(A')^2\right)-\frac{1}{2}e^{-2A(y)}\partial^2\right\}
h_{\mu\nu}(x,y)=0\, ,
$$
where $\partial^2\equiv \eta^{\mu\nu}\partial_{\mu}\partial_{\nu}$.
Introducing a trial solution in the form of a product
of an ordinary-space plane wave times a {\textit{bulk wave function}}
$h_{\mu\nu}=e^{ip\cdot x}\psi_{\mu\nu}$, we get the
{\textit{Schr\"oedinger-like}} equation
$$\left\{-\frac{1}{2}\frac{d^2}{dy^2}+
\left(A^{''}+2(A')^2\right)\right\}\psi(y)
=\frac{m^2}{2}e^{-2A(y)}\psi(y)\, . $$
We have dropped the spacetime indices and introduced the mass $m^2=-p^2$.
The existence of localized graviton in ordinary space amounts to
the existence of a normalizable localized bound state of this
equation at zero energy $m^2=0$ (zero-mode).
It is not difficult to see that indeed such a
zero mode exists. It has the wave-function
\begin{equation}
\psi_0(y)=e^{2A(y)}=
\left\{\cosh\left(\frac{\xi y}{g^2-1}
\right)\right\}^{-(g^2-1)/2}\, .
\end{equation}
In order to study the massive spectrum we must transform
the above equation into a conventional Schr\"oedinger equation.
This can be done with the help of the transformation
$y\rightarrow z=f(y)\,\,\,,\,\,\,\,\psi(y)=\Lambda \overline{\psi}$.
 Demanding the absence of first derivative terms and a standard constant
 coefficient $m^2$ in the right hand side, we arrive at
 $\Lambda =e^{A/2}\,\,\,, \,\,\,\,f'=e^{-A}$.
 The resulting Schr\"oedinger equation is
\begin{equation}
\left\{-\frac{1}{2}\frac{d^2}{dz^2}+\overline{U}(z)\right\}
\overline{\psi}(z)=m^2\overline{\psi}(z)\, ,
\end{equation}
while the potential $\overline{U}(z)$ is
\begin{equation}
\overline{U}(z)=\frac{3}{8}e^{2A}\left(2A^{''}+5(A')^2\right)
=\frac{3}{4}\left(\ddot{A}+
\frac{3}{2}(\dot{A})^2\right)\, ,
\end{equation}
and the dots denote derivatives with respect to $z$.
Note that the Schr\"oedinger equation has the form corresponding to
supersymmetric Quantum Mechanics
\begin{equation}
{\cal{Q}}^{\dagger}{\cal{Q}}\overline{\psi}=
\left\{-\frac{d}{dz}-\frac{3}{2}\dot{A}\right\}\left\{\frac{d}{dz}-
\frac{3}{2}\dot{A}\right\}\overline{\psi}=2m^2\overline{\psi}\, .
\end{equation}
This form clearly excludes the possibility of tachyonic states.
Nevertheless, in order to know whether there is a gap in the
continuum spectrum we need to know the asymptotic behaviour of the
potential.
It is possible to argue that since
$\lim_{y\rightarrow \infty}\{z\}\propto e^{\xi|y|/4}$, $
\lim_{z\rightarrow \infty}
\left\{\overline{U}(z)\right\}=\lim_{y\rightarrow \infty}
\left\{\overline{U}(z(y))\right\}
\propto e^{2A}\rightarrow 0$ and, as expected, there is no gap and
 the continuous spectrum starts from zero
energy. The situation is entirely analogous to the Randall-Sundrum
case and one could repeat the same arguments with respect to
massive graviton excitations. As a result the corrections to Newton's
law have the same, adequately suppressed, form $V(r)\propto \frac{1}{r}(1+O(r^{-2}))$.

Summarising our results, we have considered a five-dimensional
theory of a scalar coupled to gravitation with a restricted coupling to
the Standard Model. We have obtained classical Poincar\'e-invariant
solutions describing a localized smooth geometry which is everywhere finite.
 There is a limit in which the warp function of the metric tends to
 the Randall-Sundrum form and the scalar field duplicates a brane of
positive tension. The solutions depend on two parameters, a dimensionless
one $g^2$ and the dimensionfull parameter $V_0$ that is
related to Standard Model physics. For any quantum change in $V_0$
(f.e. quantum vacuum energy corrections induced by a
symmetry-breaking phase transition)
there is always a solution corresponding to the new value of $V_0$.
Thus, there is no fine-tuning involved in order to obtain a
flat four-dimensional solution. In that aspect, the solutions found display
self-tuning and resolve the four-dimensional cosmological
constant problem. However, in contrast to the recently proposed self-tuning
solutions, they have no singularities. Nevertheless, they do
involve restrictions on the bulk interactions. These restrictions, however,
are totally independent of Standard Model physics. Another crucial
 difference to the singular self-tuning solutions
 is also the fact that the
scalar field involved is not an extra bulk scalar
field but the field that forms the brane itself.

\vskip .1in

K.T. acknowledges partial support by the European Union TMR
program with contract No. ERB FMRX-CT96-0090. A. K. acknowledges
partial support by the RTN programmes HPRN-CT-2000-00122 and
HPRN-CT-2000-00131 and the $\Gamma\Gamma$ET grant E$\Lambda$/71.

\end{document}